\documentclass[acmsmall]{acmart}
\pdfoutput=1

\usepackage{amsmath}
\usepackage{graphicx}
\usepackage{hyperref}
\usepackage{mathtools} %
\usepackage{subcaption}
\usepackage{verbatim}
\usepackage{xcolor}
\usepackage{xspace}
\usepackage{ifthen}

\newcommand{\forestclaw}{{\upshape ForestClaw}\xspace}
\newcommand{\libsc}{\texttt{\upshape libsc}\xspace}
\newcommand{\pforest}{\texttt{\upshape p4est}\xspace}
\newcommand{\scda}{\texttt{\upshape\lowercase{scda}}\xspace}
\newcommand{\tetcode}{\texttt{\upshape t8code}\xspace}

\newcommand{\fscda}[1]{\textup{\texttt{scda\_}\texttt{#1}}}
\newcommand{\fxn}[1]{\textup{\texttt{#1}}}
\newcommand{\param}[1]{\textup{\texttt{#1}}}
\newcommand{\ttc}{\textup{\texttt{,}}}
\newcommand{\figlab}[1]{\label{fig:#1}}
\newcommand{\figref}[1]{Figure~\ref{fig:#1}}
\newcommand{\seclab}[1]{\label{sec:#1}}
\newcommand{\secref}[1]{Section~\ref{sec:#1}}
\newcommand{\eqnlab}[1]{\label{eqn:#1}}
\newcommand{\eqnref}[1]{\eqref{eqn:#1}}

\newcommand{\NULL}{\texttt{NULL}\xspace}
\newcommand{\fileheaderbytes}{128\xspace}

\newcommand{\textbsn}{\texttt{'\textbackslash n'}\xspace}

\newcommand{\textbsrn}{\texttt{"\textbackslash r\textbackslash n"}\xspace}
\newcommand{\textbsne}{\texttt{"\textbackslash n="}\xspace}
\newcommand{\textbsmn}{\texttt{"-\textbackslash n"}\xspace}
\newcommand{\textbsen}{\texttt{"=\textbackslash n"}\xspace}

\newcommand{\I}{\texttt{'I'}}
\newcommand{\B}{\texttt{'B'}}
\newcommand{\A}{\texttt{'A'}}
\newcommand{\V}{\texttt{'V'}}

\newcommand{\inp}{\textup{\texttt{in}}}
\newcommand{\outp}{\textup{\texttt{out}}}
\newcommand{\inoutp}{\textup{\texttt{inout}}}
\newcommand{\inoutswitch}{$\{\texttt{in}, \texttt{out}\}$}

\newenvironment{apitabular}{\begin{center}\begin{tabular}%
                {@{}p{16mm}@{}p{16mm}@{}p{94mm}@{}}}{\end{tabular}\end{center}}
\newcommand{\apireturn}[1]{\begin{apitabular}%
  \multicolumn{3}{p{126mm}}{#1}\\\end{apitabular}}
\newcommand{\apifunction}[2]{\begin{apitabular}&%
  \multicolumn{2}{p{110mm}}{\fscda{#1 (#2)}}\\\end{apitabular}}
\newcommand{\apitabreak}{\newline\hspace*{36mm}}

\newbool{draft}
\setboolean{draft}{false}

\ifthenelse{\boolean{draft}}%
{\usepackage{draftwatermark}
\SetWatermarkText{Draft}
\SetWatermarkScale{2}
\SetWatermarkColor[gray]{0.9}}%
{}%

\ifthenelse{\boolean{draft}}%
{\newcommand{\note}[1]{}}%
{\newcommand{\note}[1]{\emph{\color{purple}(Note: #1)}}}

\acmJournal{TOMS}

\setcopyright{none}
\title{\scda: A Minimal, Serial-Equivalent Format for Parallel I/O}
\author{Tim Griesbach}
\orcid{0009-0006-9217-2062}
\affiliation{%
    \institution{%
                 INS,
                 Rheinische Friedrich-Wilhelms-Universität Bonn}%
    \city{Bonn}%
    \country{Germany}}%
\email{tim.griesbach@uni-bonn.de}

\author{Carsten Burstedde}
\orcid{0000-0001-9843-1041}
\affiliation{%
    \institution{%
                 INS,
                 Rheinische Friedrich-Wilhelms-Universität Bonn}%
    \city{Bonn}%
    \country{Germany}}%
\email{burstedde@ins.uni-bonn.de}



\begin{document}

    \begin{abstract}
      \begin{center}
      \note{This is a request for comments.
            \newline
            Please email us at
            \texttt{p4est@ins.uni-bonn.de} with any remarks or suggestions.}
      \end{center}
          We specify a file-oriented data format suitable for parallel,
    partition-independent disk I/O.
    Here, a partition refers to a disjoint and ordered distribution of the
    data elements between one or more processes.
    The format is designed such that the file contents are invariant under
    linear (i.\ e., unpermuted), parallel repartition of the data prior to
    writing.
    The file contents are indistinguishable from writing in serial.
    In the same vein, the file can be read on any number of processes that
    agree on any partition of the number of elements stored.

    In addition to the format specification we propose an optional
    convention to implement transparent per-element data compression.
    The compressed data and metadata is layered inside ordinary format
    elements.
    Overall, we pay special attention to both human and machine
    readability.
    If pure ASCII data is written, or compressed data is reencoded to ASCII,
    the entire file including its header and sectioning metadata remains
    entirely in ASCII.
    If binary data is written, the metadata stays easy on the human eye.

    We refer to this format as \scda.
    Conceptually, it lies one layer below and is oblivious to the definition
    of variables, the binary representation of numbers, considerations of
    endianness, and self-describing headers, which may all be specified on
    top of \scda.
    The main purpose of the format is to abstract any parallelism
    % Thus,
    % it matches particularly well with linear space-tree codes, but is
    % in no means tied to those.
    % Quite the contrary,
    % we provide
    and provide sufficient structure as a foundation for a generic and
    flexible archival and checkpoint/restart.
    % Accordingly, we
    A documented reference implementation is available as part of
    the general-purpose \libsc free software library.

    \end{abstract}

    \maketitle

    \section{Introduction}
    \seclab{intro}

    Scientific simulations produce boundless amounts of data that is
    generally written to more or less permanent storage, such as hard disks,
    solid state memory, or magnetic tape.
    All such data is useless unless it can be read at least once.
    Two aspects stand in the way of satisfying this elementary requirement:
    \begin{enumerate}
      \item
        Large-scale simulations execute as parallel jobs that partition the
        data among multiple processes.
        Some partitions are defined by complex, indirect lookup tables,
        while others simply divide ordered data into segments.
        In either case, the amount of data files and their contents often
        depend on the number of parallel processes and/or the data partition.
        This is a purely practical limitation that makes it difficult to
        read and write from disparate jobs.
      \item
        The sheer size of the data cannot be managed by adding
        computational resources alone.
        Lossless recoding is a commonplace technique to further reduce
        the output data size.
        However, compression of data arrays as a whole often inhibits random
        and selective access to the uncompressed data and intertwines
        unfavorably with data that is partitioned in parallel.
    \end{enumerate}

    In this paper, we define the file-oriented \scda format that eliminates
    the above hindrances as much as possible.
    The format is partition-independent and thus serial-equivalent by design.
    It is intended to be generally suitable for all sorts and sizes of
    simulations by allowing the user lots of freedom in their data layout.
    This format is a container:
    It leaves the definition of variable names and attributes, binary
    encodings, content headers and the like in the hands of the user.

    Guiding principles in designing the file format have been that it shall
    be human-friendly, easy to memorize, dependency-free, and generic.
    In particular, it offers the following features.
    \begin{enumerate}
      \item
        We place all data in one big parallel file.
      \item
        The first kind of data we handle is global (unpartitioned).
      \item
        The second kind is array data of fixed or variable size per element.
      \item
        The file contents do not depend on the job size and its data
        partition.
      \item
        The format is easily readable by humans, primitive scripts, and
        complex programs alike.
    \end{enumerate}

    Writing one file is naturally the most straightforward way to keep
    the format serial equivalent.
    In practice, this approach is well supported by the MPI
    standard \cite{Forum97},
    and the file contents can be read and written efficiently in parallel.
    We enable selective random data access even with variable-size
    array elements and/or per-element compression.
    Incidentally, having one file independent of the job size greatly
    simplifies downstream file management for archival and
    checkpoint/restart.

    In practice, even assuming that a parallel file system and a fast MPI
    I/O implementation are available, supporting general numerical
    simulations is non-trivial due to the mesh data to be stored.
    For the special case of contiguous indexed partitions, such as
    (but not limited to) those arising from space-filling-curve partitions,
    the mesh data is but a special case of array data that may be easily
    stored in the same file as the numerical information.
    Arbitrary mesh data can be stored using variable-size and -length arrays
    as long as its numbering is global and serial-equivalent.

    It would go beyond the scope of this article to reference all
    parallel file formats to completion.
    Instead, we will briefly relate to three especially relevant ones.
    \begin{enumerate}
      \item The VTU appended binary format \cite{Kitware10}
            is primarily intended to write
            the mesh connectivity of a numerical simulation together with
            associated numerical data.
            It consists of an XML header that contains multiple arrays of
            mesh metadata, such as elements' types and their vertices.
            After the header, the data is written as flattened binary
            arrays.
            This format is well suited for single-file partition-independent
            graphics output since both the header and the data may be
            written in parallel using the MPI I/O standard.
            The \forestclaw code \cite{BursteddeCalhounMandliEtAl14}
            does it this way.
      \item HDF5 \cite{HDF23, HDF19} is a hierarchical data format that
            provides a naming scheme akin to a file system and encodes
            types, attributes, and objects. The format is so general
            that it practically requires to link to a heavy software
            library to process the files.
            Our goals are quite different in that the \scda format
            specification is short, and access functions are easily
            programmed by the interested reader.
            We can only achieve such simplicity by forgoing any ambition to
            implement types, or associations.
            The best of both worlds may be to write an HDF5 file of global
            parameters to memory, to save that as an \scda block
            section, and to append partitioned data as native \scda arrays.
      \item The NetCDF format \cite{RewDavisEmmersonEtAl97}
            is machine-independent and well supported by software that
            ensures backward compatibility and handles various common issues
            like endianness.
            Parallel access to this format has been added somewhat later
            \cite{LiLiaoChoudharyEtAl03}.
            More recently, NetCDF has been integrated with an optional HDF5
            backend, which makes the format more flexible but also
            increasingly dependent on linking to third-party libraries.
            In comparison, \scda is transparent and inherently scalable
            but leaves the choice of binary data conventions entirely to the
            user.
    \end{enumerate}

    The \scda format support is naturally implemented using MPI I/O, which
    usually addresses a parallel file system such as Lustre
    \cite{BraamSchwan02}.
    It can be said that the \scda format itself is one level below the
    typical functionality of a HDF5 or NetCDF specification.
    While the latter natively support types and strided and offset data,
    such can be encoded in \scda by writing user-defined lookup tables.
    We leave indirect addressing and encoding of data variables to the
    user as application developer.

    The HDF5 format supports transparent compression, which is a necessity
    to work with large scale simulation data.
    The LightAMR format \cite{StrafellaChapon22} targets
    distributed-parallel cell-based adaptive meshes using lossless data
    compression based on concrete assumptions on the mesh data structure.
    In contrast, our aim is to be generic without loosing simplicity, and
    therefore we assume nothing but a contiguous indexed partition.
    Another example for an application-specific parallel checkpoint/restart
    focused on adaptive mesh data is described in
    \cite{JainWeideChawdharyEtAl21}.
    We refer to \cite{BoitoZanonInacioEtAl18} for a survey on
    parallel I/O for HPC
    including a general introduction on the topic.

    Our approach is to define the \scda format oblivious to the data
    contents and to support compression by a convenience layer on top of it.
    This approach allows us to implement transparent access and
    deflate/inflate, both for global objects and array data, as an optional
    convention.
    Stacking another convention for encryption would be relatively simple.
    We encode array data per element, which has the downside to include
    more overhead than monolithic compression of a whole array.
    The upside is that parallel array access remains fast and inherently
    scalable.

    The \scda format specification is laid out in \secref{fileformat}.
    The proposed convention for transparent data compression
    is described in \secref{compression}.
    These two sections underline \scda as a data format
    independent of any particular software or hardware.
    Still, to allow for testing and demonstration and to jump start the
    use of \scda in third-party scientific software, we provide several
    example use cases in \secref{application}.
    These rely on a reference implementation of \scda newly added to the
    \libsc software library.
    A minimal C API, its documentation, and the rationale in calling its
    entry points are described in the appendix of this paper.

    \section{The \scda format specification}
    \seclab{fileformat}

    In this section, we cover the \scda file format in its entirety.
    It is understood as an unambiguous specification of a data
    layout that is independent of the software used to write it.
    Every byte of the file written is well defined by the user's input data,
which is treated as a sequence of raw bytes without regard to multibyte
characters or alternative encodings, NUL termination, escapes, binary
representation of numbers, etc.
    On reading, the original input data can be reproduced exactly from the
file's contents.

    The file consists of one or more sections without gaps before or after.
    The first section is always the file header that includes magic bytes
    and version defined by the file format and then a vendor and a user string:
    \begin{description}
        \item[F] The file header section comes first.
    \end{description}
    The file header section must not occur again.
    The remainder of the file consists of zero or more data sections, each
of which must be of one of the following four types:
    \begin{description}
        \item[I] an inline data section,
        \item[B] a data block of a given size,
        \item[A] an array of given length and fixed element size,
        \item[V] an array of given length and variable element size.
    \end{description}
    The four data section types are of ascending generality.
    This means, for example, that inline data may also be written as a
suitably defined block, a block may be written as a suitable 1-element
array, or that a fixed-size array may also be written as a suitably defined
variable-size array, at the expense of increased redundancy and file size.
    Each section contains a user string as well as the count and size
information necessary for its respective purpose, which are all considered
part of the definition of the input data.
The sections are composed of a small selection of parameterized entries,
namely
    \begin{itemize}
        \item
      the file format magic and version (8 bytes),
        \item
      a vendor string (24 bytes),
        \item
      a section type and user string (64 bytes),
        \item
      a non-negative integer variable (32 bytes),
        \item
      data bytes.
    \end{itemize}

    In practice, the strings in the file structure are often input by the
user as a proper C string, employing escapes and avoiding
non-printable characters and NUL, but the format specification simply demands a
sequence of bytes whose length is limited by an explicitly defined maximum.
    The format allows for byte counts of data elements and array lengths
requiring up to 26 decimal digits.

    To align the sections as well as the entries of each section at
sensible power-of-2 byte boundaries, and to provide human-friendly line
breaks, we introduce two types of padding, one for strings and counts using
the \texttt{'-'} character and one for data bytes using \texttt{'='}.
    The first kind allows to infer the original byte count from the padding,
while the second kind relies on an input byte count known by construction.
The padding bytes are included in the above list of entries.

    \subsection{Padding}
    \seclab{padding}

    Padding means to add bytes to input information, which may itself be of
length 0, to the right according to a well-defined rule.
    The first goal is to ensure that the byte count of the padded data is
divisible by a specified divisor, which renders certain section entries
constant in byte length and generally simplifies the layout of the file.
    The second goal is to render the experience of a human opening a
conforming file in a text editor as pleasant and consistent as possible.
    We achieve this predominantly by adding line breaks in selected places.
    The type of line break written may be chosen by the user to MIME or
Unix.  On reading the file, this choice (or lack of it) has no effect.

    \subsubsection{Padding strings and counts to a fixed number of bytes}
    \seclab{pad_user_string}

    User strings are of variable length, and byte and element counts require
a variable number of decimal digits.
    In this situation, we require right padding to extend a byte sequence of
length $0 \le n \le d - 4$ to length $d$.
    The number of padding bytes is thus $p = d - n \ge 4$, and we define
\begin{equation}
    \eqnlab{userpadding}
    \texttt{padding} (\text{'\texttt{-}' to $d$})
    =
    \texttt{'\textvisiblespace'},
    (p - 3) \times \texttt{'-'},
    q.
\end{equation}
    Here \texttt{\textvisiblespace} denotes the ASCII space \#32, and the
dash is the ASCII dash \#45.
    The rightmost part of this padding, $q$, refers to two arbitrary bytes,
which must be \textbsmn for Unix and \textbsrn for MIME.
    The C-style escapes denote the carriage return or line feed byte.
    Since we define $d$ in the format, the padded data can be parsed from
the right to infer $p$ and hence $n$, which then allows to read the input
sans padding from the left.

    \subsubsection{Padding of data bytes}
    \seclab{pad_data_block}

    We pad input data bytes with the aim to make the number of bytes written
divisible by $D$ (which, for the purpose of this format, is always 32).
    The number of padding bytes $p$ is at least $7$ and at most $D + 6$,
defined as the unique integer in this range that makes $n + p$ evenly
divisible by $D$.
    In contrast to the previous section, we choose the contents of the
padding depending on the last byte of the input data (if any).
    This is to acknowledge that the input is often ASCII armored by the
user, with or without a terminating line break.
    For visual consistency, we prefer to pad with a line break only if the
data does not end in one.
    Thus, we define
\begin{equation}
    \eqnlab{datapadding}
    \texttt{padding} (\text{'\texttt{=}' mod $D$})
    =
    P, Q \times \texttt{'='}, R.
\end{equation}
    If $n > 0$ and the last input byte is \textbsn, we set
$P$ to \texttt{"=="}.
    Otherwise, it is \textbsrn for MIME and \textbsne for Unix.
    The symbols $Q$ and $R$ are defined in Table~\ref{tab:defineQR}.
    By construction, the data padding is always $p$ bytes long.
\begin{table}
   \begin{center}
     \begin{tabular}{lll}
       \toprule
                 & \multicolumn{1}{c}{$Q$} & \multicolumn{1}{c}{$R$} \\
       \midrule
       MIME      & $p - 6$ &
  \texttt{"\textbackslash r\textbackslash n\textbackslash r\textbackslash n"} \\
       Unix      & $p - 4$ &
  \texttt{"\textbackslash n\textbackslash n"} \\
       \bottomrule
     \end{tabular}%
   \end{center}%
   \caption{Variables for the padding of data bytes \eqnref{datapadding}.
            The double quoted strings are understood as escaped characters
            in C syntax without the terminating NUL.}
   \label{tab:defineQR}%
\end{table}%
If neither MIME nor Unix line endings are desired, the data padding may
consist of $p$ arbitrary bytes.
The file format is defined such that the number of data padding bytes can
always be inferred from the preceding file contents, and the padding
bytes are ignored on reading.

    \subsection{File header section}
    \seclab{fileheader}

    \begin{figure}
        \centering%
        \includegraphics[width=0.9\textwidth]{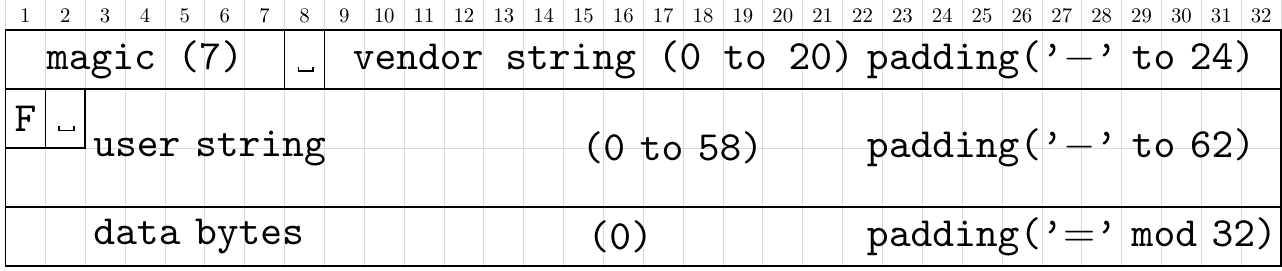}%
        \caption{The file header \textbf{F} is \fileheaderbytes bytes long,
                 displayed here with 32 bytes per row.
    We denote the byte length $n$ of an entry in parantheses.
    The magic encodes a format identifier and version as
    \texttt{sc\%02xt\%02x} in printf notation.
    The identifier for \scda is $(\mathrm{da})_{16} = 208$.
    The format version counts in hexadecimal from the present \texttt{scdata0}
    ($(\mathrm{a0})_{16} = 160$) to \texttt{scdatff}
    ($(\mathrm{ff})_{16} = 255$), offering a range of 96 values.
         The vendor string is hardcoded by an individual software
implementation, and the user string is arbitrary input.
We write zero data bytes to prompt consistent padding.
         The padding bytes are defined in \secref{padding}
and conclude with a blank line.%
}%
        \Description[Byte layout of the file header section]{%
    The pictures shows a graphical layout of the bytes that
    constitute the file header section.
    The vendor string as at most 20 bytes long and '-'padded to 24.
    The user string is at most 58 bytes long,
    preceded by the letter F and a space, and - padded to 62 bytes.
    The zero data bytes are = padded to 32 bytes.}%
        \figlab{file_header}%
    \end{figure}

    Every file begins with a header section that encodes the file format version
and offers limited implementation- and user-specific string data (where, as
explained above, we do not interpret strings and generally allow for a
sequence of arbitrary raw bytes).
Like every other section, it identifies itself with a specific letter,
\textbf{F} in this case.
The file header allows for a vendor string of at most 20 bytes and a user
string of at most 58; please see \figref{file_header} for details.
If more context data is required, it can be written using separate inline
and block sections as described below.

We formally represent the file header as a function of the version number
$v$ and its entries,
    \begin{equation}
        \texttt{F} (v, \texttt{vendor string}, \texttt{user string}).
    \end{equation}
    To simplify notation, we understand the lengths of the vendor and user
strings to be part of the input.

    \subsection{Inline data section}
    \seclab{inline}

    The inline data section is intended to write a small data item to the file,
such as a single configuration or status variable or a short comment, or a
record of binary data.
It requires the user to input exactly 32 bytes of data.
These may include any kind of user-defined structuring or padding to shape
the visual appearance of the file.
In particular, this section type enables the user to
style an arbitrary prefix to a following section since it does not
enforce a trailing blank line.
The inline section comes with a user string as all others, and always
has a size of 96 bytes.

    \begin{figure}
        \centering%
        \includegraphics[width=0.9\textwidth]{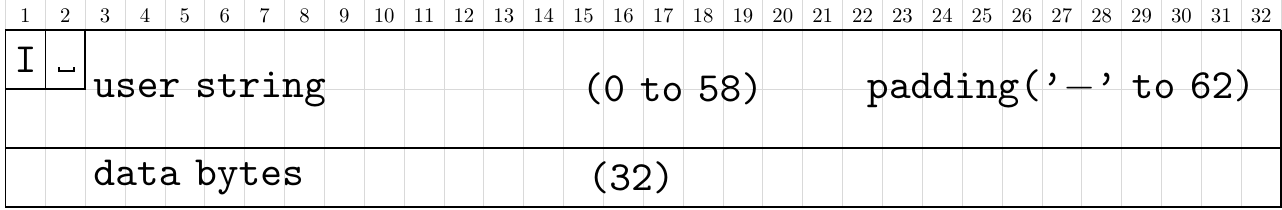}%
        \caption{The format of the inline data file section \textbf{I}
using the notation introduced below \figref{file_header}.
It is important to note that the inline type is the only one with unpadded
data.
The input data therefore must amount to exactly 32 bytes.
This exception grants maximum freedom to the user to visually arrange a
collection of individual data items in the file.
The user may place their own structuring and padding within this space.}
        \figlab{inline}
    \end{figure}
    The detailed structure of the \textbf{I} data section is depicted in
\figref{inline}.
We formally represent inline data as a function of its entries,
    \begin{equation}
        \texttt{I} (\texttt{user string}, \texttt{data bytes}).
    \end{equation}

    \subsection{Data block of given size}
    \seclab{block}

    The data block file section has the purpose to store global
unpartitioned data.
    The byte length of the block is arbitrary as long as it fits into a
26-digit decimal number.
In addition to the usual user string, the length must be provided along with
the input data.

    The block section is the first type that contains a size parameter,
    in this case the input length $E$.
    The format encodes this number as the letter \texttt{'E'}, a space
character, and a non-negative integer of at most 26 decimal digits.
    The remainder of the size entry is padded to 32 bytes.

    This block section type \textbf{B} can be used for any kind of global
data, e.\ g.\ a global simulation context or some
    user-defined metadata that is used to interpret the rest of the file.
The detailed structure of the block section type is depicted in
\figref{block}.
    We formally represent a block section as a function of its
entries to later refer to this type, namely
    \begin{equation}
        \texttt{B} (\texttt{user string}, E, \texttt{data bytes}).
    \end{equation}

    \begin{figure}
        \centering%
        \includegraphics[width=0.9\textwidth]{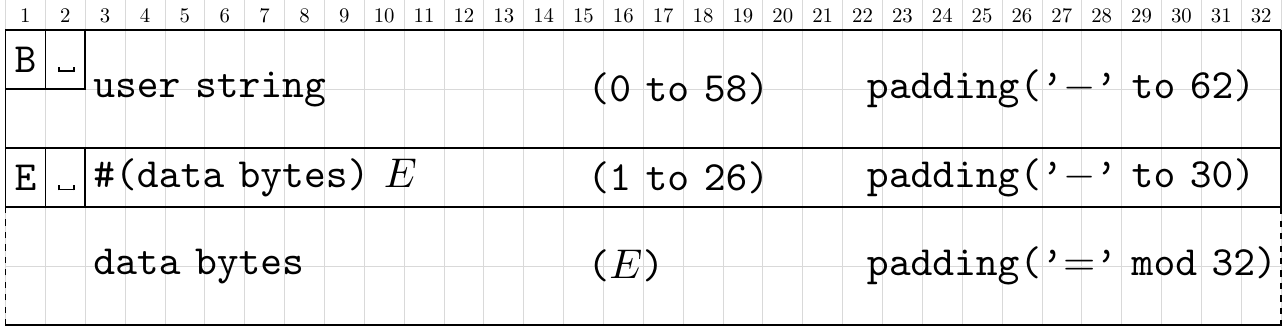}%
        \caption{The format of the block data file section \textbf{B}
displayed with 32 bytes per row.
        The number of data bytes $E$ is printed in decimal without leading
spaces or zeros.
        The dashed vertical lines indicate that the data bytes and their padding
        may consume an arbitrary amount of 32-byte lines in the gridded display.}
        \figlab{block}
    \end{figure}

    \subsection{Array of fixed-size elements}
    \seclab{fixed}

    The fixed-size array is the simplest file section that enables the user
to read and write data in parallel.
    Its purpose is to store an array of a given element count and a fixed
element byte size.
    The number of elements $N$ and the data size per element $E$ must be
within the usual limit of a 26-digit decimal number.
    We write each as a number entry as introduced above with the block section.

    This file section type can be used to store uniform
numerical data, mesh data or mesh-associated data, especially if a numerical
application distributes mesh elements and associated data in parallel.
    This file section type allows for efficient parallel I/O.
    Its detailed structure is depicted in \figref{fixed}.

    We formally represent a fixed-size data array \textbf{A} as a
function of its entries,
    \begin{equation}
        \texttt{A} (\texttt{user string}, N, E, \texttt{data bytes}).
    \end{equation}

    \begin{figure}%
        \centering%
        \includegraphics[width=0.9\textwidth]{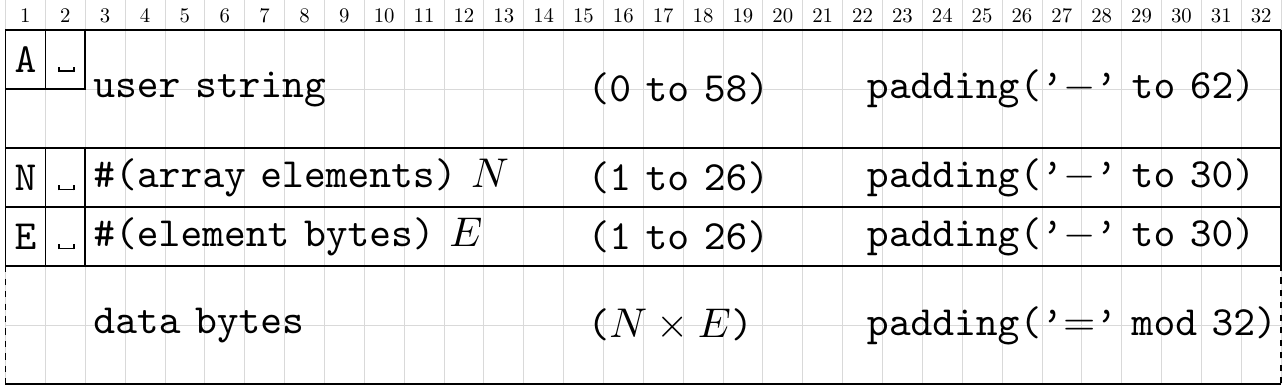}%
        \caption{The format of the fixed-size array section \textbf{A}.
            The number of array elements $N$ and the number of bytes per
element $E$ are printed as decimal integers as in previous sections.
$E$ is the element byte size for each of the $N$ array elements, since
            the array element size is equal for all elements.
            The data bytes contain the array data concatenated over the
elements and only padded once after the last element.}%
        \figlab{fixed}%
    \end{figure}%

    \subsection{Array of variable-size elements}
    \seclab{variable}

    The most general file section type we provide allows to store an array
of data elements with varying sizes.
    To encode this information, we write the number of array elements and then
one number entry for each element before we add the concatenated array contents.

    The structure of the this section coincides with that of the fixed-size
array up to and including the number of elements $N$.
    This part is succeeded by a list of number entries that store the byte
size of each array element.

    The variable element-size array can be used e.\ g.\ to store hybrid
meshes since such a mesh may induce varying data sizes
    depending on each mesh element's shape.
    The data of $hp$-adaptive element methods is a prime example requiring
this section type.
Another application example is writing mesh-oriented data that is
    compressed per element, which usually ensues variable compressed data sizes
among the elements.
    In fact, this is one use case we propose below in \secref{compression}.
    The detailed structure of the variable-size section type \textbf{V} is
depicted in \figref{variable}.
    \begin{figure}%
        \centering%
        \includegraphics[width=0.9\textwidth]{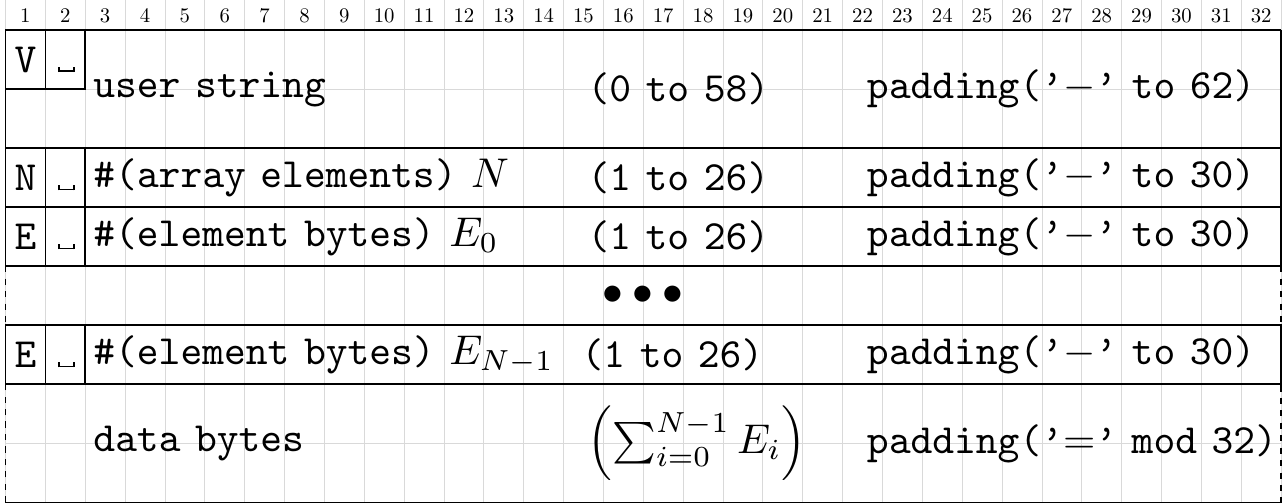}%
        \caption{The format of the variable-size array section \textbf{V}.
            The number of data bytes $N$ and the number of element bytes
$E_i$ for $i\in\{0,\ldots,N-1\}$ are encoded using at most 26 decimals digits each.
The dashed vertical lines indicate a variable number of 32-byte chunks in the file.
Each entry $E_i$ is the element size in number of bytes of the $i$-th array
element.
The data bytes arise from concatenating the array's element data in order
and padding only once after the last element.}
        \figlab{variable}%
    \end{figure}%

    We formally represent an array of variable-size elements as a function
of its entries, in this case
    \begin{equation}
        \texttt{V} (\texttt{user string}, N, (E_i)_{i\in\{0, \ldots, N-1\}},
                    \texttt{data bytes}).
    \end{equation}

    \section{Per-element data compression}
    \seclab{compression}

    Lossless compression is useful to reduce the size of files written to
    disk.
    It is offered by many image formats such as PNG \cite{AdlerBoutellBowlerEtAl03}
    and scientific graphics and data formats like VTK and HDF5.
    The compression is generally transparent, meaning that no knowledge
    of the compression algorithm is required on the user side.
    Since we intend to keep the \scda format truly minimal, we do not
    include compression in the specification laid out in \secref{fileformat}.
    Instead, we
    % use this \secref{compression} to
    suggest an additional convention to store compressed data
    using the basic types of the \scda format as wrappers.

    Compressing a data block changes its size such that the data size $E$ encoded
    in an \scda block section takes the meaning of the compressed size.
    Similarly, both fixed and variable size arrays yield compressed data
    of variable element size.
    The uncompressed sizes must then be written as additional metadata.
    We handle this requirement by encoding each compressed file section
    using two of the raw section types \textbf{I}, \textbf{B}, \textbf{A},
    and \textbf{V}.
    This approach lets us write a second user string, which we repurpose to
    identify the compression convention and version of the algorithm.
    If the type of the first raw section and its user string match as
    listed below for \eqnref{compress_block}, \eqnref{compress_fixed} or
    \eqnref{compress_variable}, the remainder of the two raw sections must
    fully conform to the convention to prevent an error on reading.
    %
    % that is read as compressed file section must
    % conform to the compression convention if the first file section
    % type and the user string of the respective file section are as in
    % one of the first file sections in

    \subsection{Compression algorithm}
    \seclab{compressalgo}

    We compress the input data for a block, or for each data element of an
    array, on its own by the same elementary algorithm.
    In the first of its two stages, the data is transformed into the
    following items concatenated:
    \begin{enumerate}
      \item The uncompressed size written as 8-byte unsigned integer in
            big-endian (MSB first).
      \item The byte \texttt{'z'}.
      \item The data as an RFC 1950/1951 deflate stream \cite{Deutsch96}
            using any legal compression level.
    \end{enumerate}
    We recommend zlib's best compression and the
    \texttt{compress2} function \cite{GaillyAdler23},
    but it is possible to conform by using level 0 (no compression),
    which is easy to hardcode if zlib is not available.

    In the second stage, the output of the first is base64 encoded
    to lines of 76 code bytes and 2 bytes for a general line break.
    These latter two bytes are arbitrary, but must be \textbsrn for the
    MIME style and \textbsen for the Unix style.
    The same two bytes are added after the last line of
    encoding if it is short of 76 bytes.
    We refer to the byte length of the resulting stream as the
    compressed size.
    The result is in ASCII (as long as the line breaks are) and
    optionally broken into lines of acceptable length.
    As such it is written into the elementary data section types as
    defined in \secref{fileformat}.

    On reading the compressed data, we exploit the fact that its length is
    known by file context.
    Thus the data is base64 decoded and its uncompressed size is extracted
    from the first 8 bytes of the result.
    This information suffices to allocate memory as necessary and to execute
    zlib's \texttt{uncompress} function \cite{GaillyAdler23}
    starting at the tenth byte to recover the original input.

    We notice that three redundant checks are involved in reading the data:
    The Adler32 checksum \cite{DeutschGailly96} executed inside zlib,
    comparing the
    uncompressed size with the result of decompression, and verifying that
    the ninth byte of the decoded base64 data is indeed \texttt{'z'}.

    \subsection{Compression of a data block}
    \seclab{compressblock}

    To write a compressed data block, we require one metadata item, namely
    its uncompressed size.
    We write this into an inline section (see \secref{inline}) using its 32
    data bytes as pictured in \figref{compressioninline}.
    \begin{figure}
        \centering%
        \includegraphics[width=0.9\textwidth]{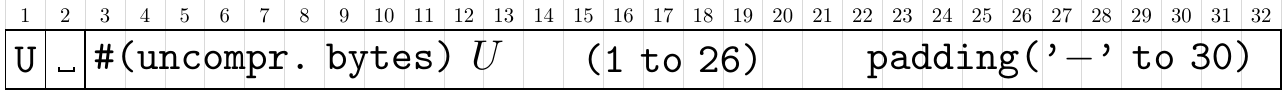}%
        \caption{The data of an inline data section can be used to encode
                 the uncompressed size of a data block or of one element of an array.
                 We mimic the number convention for the $N$ and $E$ entries of the
                 \scda format and use the same padding specification; see
                 \secref{fileformat}.}
        \Description{...} %
        \figlab{compressioninline}
    \end{figure}
    The user string of the inline section is set to a magic string that
    identifies the \scda compression convention and its version
    $(\mathrm{00})_{16}$.
    The inline data section is succeeded immediately by a data block
    storing the original user string and the data compressed by the
    algorithm described in \secref{compressalgo}.
    In symbols, the compressed data block is written as
    \begin{equation}
    \begin{aligned}
        \texttt{I} (&"\texttt{B compressed \scda 00}",\\
        &\texttt{data bytes} \text{ as in \figref{compressioninline} with
         $U = \#(\texttt{uncompressed data bytes})$}),\\
        \texttt{B} (&\texttt{user string},\\
        &N = \#(\texttt{compressed data bytes}), \texttt{compressed data bytes}).
    \end{aligned}
    \eqnlab{compress_block}
    \end{equation}

    \subsection{Compression of an array of fixed element size}
    \seclab{compressfixed}

    A fixed-size array has the same uncompressed data size for every
    element.
    As above for the data block, it suffices to store this one number in a
    prepended inline data section, this time with a magic user string
    that contains the letter \texttt{'A'} instead of \texttt{'B'}.

    Now, we must use a variable-size array to store the array with
    compressed elements that generally differ in size.
    This array has the same number of array elements $N$ as the uncompressed
    fixed-size array and stores the compressed sizes per element together
    with the data bytes compressed per-element.
    These considerations lead to the following format for fixed-size array data:
    \begin{equation}
    \begin{aligned}
        \texttt{I} (&"\texttt{A compressed \scda 00}",\\
        &\texttt{data bytes} \text{ as in \figref{compressioninline} with
         $U = \#(\texttt{uncompressed element bytes})$}),\\
        \texttt{V} (&\texttt{user string},
         N = \#(\texttt{array elements}),\\
        &(E_i=\#(\texttt{$i$-th compressed data bytes}))_{i\in\{0,\ldots,N-1\}},
         (\texttt{compressed data bytes})_i).
    \end{aligned}
    \eqnlab{compress_fixed}
    \end{equation}

    \subsection{Compression of an array of variable element size}
    \seclab{compressvariable}

    In contrast to the compression of the fixed-size elements array, the
    variable-size version introduces individual uncompressed element sizes.
    We record these using a fixed-size array section, which is more general
    than the inline section type used for a compressed block or array of
    fixed element size.
    We write each element of the uncompressed sizes array as a 32-byte
    entry as defined in \figref{compressionv}.
    \begin{figure}
        \centering%
        \includegraphics[width=0.9\textwidth]{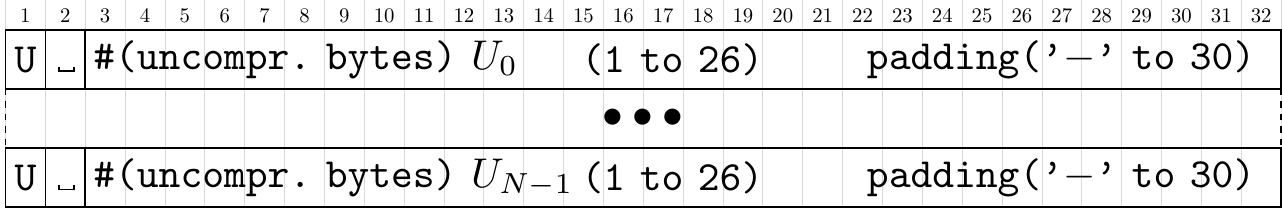}%
        \caption{We display the data bytes of a fixed size array of $N$
elements used to encode a list of uncompressed sizes with 32 bytes each.
The encoding is the same as in \figref{compressioninline}.
We use this convention to write the metadata for a compressed array of
variable element byte size.}%
        \Description{...} %
        \figlab{compressionv}%
    \end{figure}%

    Immediately following, the compressed data is written as an array of
    variable-size elements exactly as described for the compression of
    fixed-size element data.
    To sum it up, our convention for a compressed array of variable-size
    elements is
    \begin{equation}
    \begin{aligned}
        \texttt{A} (&"\texttt{V compressed \scda 00}",\\
        &N = \#(\texttt{array elements}), E = 32,
         \texttt{data bytes} \text{ as in \figref{compressionv}}),\\
        \texttt{V} (&\texttt{user string}, N = \#(\texttt{array elements}),\\
        &(E_i=\#(\texttt{$i$-th compressed data bytes}))_{i\in\{0,\ldots,N-1\}},
         (\texttt{compressed data bytes})_i).
    \end{aligned}
    \eqnlab{compress_variable}
    \end{equation}

    \section{Application examples}
    \seclab{application}

    \note{To discuss the specification with peers,
          the application results are not yet required.}

    \section{Conclusion}

    \note{To discuss the specification with peers,
          the conclusion is not yet required.}

    \section*{Acknowledgement}

    This work is supported by the Bonn International Graduate School for
    Mathematics (BIGS), as well as travel funds, by the Hausdorff Center for
    Mathematics (HCM) at the University of Bonn funded by the German
    Research Foundation (DFG) under Germany's excellence initiative EXC 59
    -- 241002279 (Mathematics: Foundations, Models, Applications).

    We gratefully acknowledge partial support under DARPA Cooperative Agreement
    HR00112120003 via a subcontract with Embry-Riddle Aeronautical University.
    This work is approved for public release; distribution is unlimited.
    The information in this document does not necessarily reflect the
    position or the policy of the US Government.

    \appendix

    \section{Authors' reference implementation}
    \seclab{workflow}

    The \scda file format specification together with the optional
compression and encoding conventions introduced in the main text are defined
without dependence on a specific implementation or the parallel partition of
any given job.
    In practice however, the partition determines which process governs
which part of the data, and in consequence which windows onto the file it
accesses.
    Therefore, we propose an exemplary functional interface that explicitly
reflects both the partition and the data to be written and read.

    We accompany the interface with a documented reference implementation
as part of the general-purpose \libsc free
software library \cite{BursteddeWilcoxIsaac23}.
    \libsc is used in particular from the adaptive mesh refinement software
libraries \pforest \cite{BursteddeWilcoxGhattas11, Burstedde23a} and \tetcode
\cite{BursteddeHolke16, HolkeBursteddeKnappEtAl23}, which are themselves
integrated with a variety of HPC applications and libraries.

    \subsection{The parallel partition}
    \seclab{partition}

    To begin, we detail the encoding of the partition for the parallel
writing and reading of distributed array data.
    The fundamental assumption is that each array element is assigned to
    precisely one process and that this assignment is monotonous by rank.
    The data for each array element is written by its owner.
    The partition on reading, on the other hand, can be defined afresh for
    each array.

    In addition to a given data array, we define various count/offset arrays with
elements $A_j \in \mathbb{N}_0$ using the expression $(A_j)_{J'\le j<J}$ for
the contiguous range from $J'$ to $J$ exclusive.
$J' \le j$ may be omitted if $J' = 0$.
    The partition of a data array with $N$ global elements among $P$ processes,
    for instance, is expressed
    as $(N_p)_{<P}$,
    where $0 \le N_p \le N$ are the per-process counts, $0 \le p < P$,
    with offsets
    \begin{equation}
        \eqnlab{partitioncumulative}
        C_p = \sum_{q = 0}^{< p}N_q \quad \Longrightarrow \quad
        C_0 = 0, \quad C_P = N
        .
    \end{equation}
    We expose the byte sizes per element as $(E_i)_{< N}$ for a
total data size $S$ and per-process byte sizes
    \begin{equation}
        \eqnlab{sp}
        S_p = \sum_{i = C_{p}}^{< C_{p + 1}} E_i
        \quad \Longrightarrow \quad
        S = \sum_{p = 0}^{<P} S_p .
    \end{equation}

    For the special case of a fixed-size array
    the element size is a constant $E$,
    \begin{equation}
      \eqnlab{spfixed}
      E_i = E \quad \Longrightarrow \quad
      S_p = N_p E \quad\text{and}\quad
        S = N E
      .
    \end{equation}

    \subsection{Parameter conventions}
    \seclab{conventions}

    We proceed outlining various parameters and return values that appear
    repeatedly.
    The following descriptions depend
    on whether the respective function is a writing
    or a reading function.

        \paragraph{Parameters with multiple appearances}
        \begin{apitabular}
            \param{f} & \inoutp & Opaque file context $\{\text{created}, \text{opened}\}$ by
\fscda{fopen} with mode $\{\texttt{'w'}, \texttt{'r'}\}$.\\
            \param{userstr} & \inoutswitch & The user string that is $\{\text{written to},
\text{read from}\}$ a section header.
            The format permits up to 58 bytes of arbitrary data, which we pass as a
            character array of appropriate length plus a NUL for safety.
\begin{comment}
            While the format permits arbitrary binary data as string contents, which we
            return faithfully on reading, the API expects a NUL-terminated string on writing.
\end{comment}
\\
            \param{root} & \inp & The $\{\text{writing}, \text{reading}\}$ process on which
a non-partitioned data item is $\{\text{present}, \text{allocated}\}$.\\
            \param{indirect} & \inp & A Boolean to determine whether a data array
            is indirectly addressed.  Indirect addressing requires passing
an array of pointers to the element data items as opposed to passing
the array data as one contiguous memory range.\\
            \param{encode} & \inp & A Boolean to specify whether a file section
             is written according to the compression convention introduced in
             \secref{compression}.\\
            \param{err} & \outp & An error code that is set from every function call;
                          see \secref{errorman}.\\
        \end{apitabular}
        \paragraph{Return of all $\{\text{writing},
                   \text{reading}\}$ functions}
        \apireturn{%
        The file context \param{f} passed on input,
        which is updated to continue $\{\text{writing}, \text{reading}\}$
using any function in Section~$\{\text{\ref{sec:writing}}, \text{\ref{sec:reading}}\}$
        and to eventually close the file using \fscda{fclose}.
        On error, the file is closed as is, the file context is deallocated, and \NULL
is returned.}

    Moreover, all function parameters above and in the remainder of this
    documentation are collective except for \param{dbytes} and \param{$(E_i)$}.
    For collective input parameters, it is an unchecked runtime error if
they are indeed not collective, and the behavior is undefined.
    For output variables, a collective parameter is filled identically on
all processes.

%    In \secref{fileformat} user strings were introduced as arbitrary
%    binary data but for our API we restrict ourselves for the
%    sake of simplicity to C strings, i.e.\ user strings only
%    contain characters.

    \subsection{Open and close}
    \seclab{openclose}

    With this API, all workflows start with collectively opening a
file and end with collectively closing the file.
    We do not supply a mechanism to append to existing files, and the only
possibility to write to a file is to create a new one or to overwrite an
existing one.

Therefore,
    all writing workflows start with the function \fscda{fopen} with
\texttt{'w'} as mode.
    For the case of reading we provide \fscda{fopen} with \texttt{'r'}
as mode.
    This means we use the semantics of \fxn{fopen} \cite{ISO90}
without any further mode modifiers.
In the following, we give a brief description of its call convention
relying on some general parameters found in \secref{conventions}.

        \subsubsection{Open a file}

        \apifunction{fopen}{mpicomm, filename, mode, userstr, err}

        \paragraph{Parameters}
        \begin{apitabular}
            \param{mpicomm} & \inp & The MPI communicator is used to
            collectively open the file.\\
            \param{filename} & \inp & The path to the file that is intended
            to be opened.\\
            \param{mode} & \inp &  Either \texttt{'w'} for writing to a newly created file
            or \texttt{'r'} to read from an existing file.\\
        \end{apitabular}

        \paragraph{Return}
        \apireturn{A pointer to an allocated file context that can be used by the functions
        introduced in \secref{writing} if \param{mode} is set to \texttt{'w'} or
        for \param{mode} set to \texttt{'r'} by the functions introduced in
        \secref{reading}.
        In a case of error the function returns \NULL and the error code can be
        examined for details.}
    The opaque file context maintains a file cursor that only moves forward.
    All function calls on a file context advance this file cursor by one section,
    which in the case of opening is the file header (see \secref{fileheader}).
    Eventually, the API workflow is terminated by collectively closing the
    file, deallocating the file context using \fscda{fclose}.

        \subsubsection{Close a file}

        \apifunction{fclose}{\param{f}\ttc \param{err}}

        \paragraph{Return}
        \apireturn{0 if and only if the function is successful.
            The file context is deallocated regardless.}

    \subsection{Writing}
    \seclab{writing}

    This section covers the API of the writing functions. We introduce one writing function per
    file section type as introduced in \secref{fileformat}.  As announced before, we
    use the parameter conventions introduced in \secref{conventions}.
    Our implementation writes Unix line breaks.

        \subsubsection{Write an inline section}

    Writing an inline data section as motivated in
    \secref{inline} follows the semantics of
    \fxn{MPI\_Bcast} \cite{Forum97}.
    As with all other functions, the call is collective over the file parameter \param{f}.

        \apifunction{fwrite\_inline}{f, dbytes, userstr, root, err}

        \paragraph{Parameters}
        \begin{apitabular}
            \param{dbytes} & \inp & On the \param{root} process exactly 32 bytes. Ignored
            on all other processes.
        \end{apitabular}

        \subsubsection{Write a block section}

	The function \fscda{fwrite\_block} follows the semantics
        of \fxn{MPI\_Bcast} even more closely since the block has a
        user-defined data size (cf.\ \secref{block}).

        \apifunction{fwrite\_block}{f, dbytes, $E$, userstr, root, encode, err}

        \paragraph{Parameters}
        \begin{apitabular}
            \param{dbytes} & \inp & On the \param{root} process exactly
                                    $E$ bytes. Ignored on all other processes.\\
             $E$ & \inp & The number of block data bytes written by the
             \param{root} process.
        \end{apitabular}

        \subsubsection{Write an array of same size elements}

	The simplest
	function to write distributed data in parallel is \fscda{fwrite\_array}.
	It writes an array of fixed-size elements (cf.\ \secref{fixed}).
        This function follows the semantics of \fxn{MPI\_Allgather}
\cite{Forum97} in the sense that the receive buffer is the file and the send
buffer the data bytes local to the calling process $p$.

        \apifunction{fwrite\_array}{f, dbytes, $(N_q)_{<P}$, $E$,
                     \apitabreak
                     indirect, userstr, encode, err}

        \paragraph{Parameters}
		\begin{apitabular}
            \param{dbytes} & \inp
            & On a respective process $p$, the local $N_p$ array elements
with $E$ bytes per element addressed according to the \param{indirect}
parameter.\\
            $(N_q)$ & \inp
            & The array of elements per process
            defining the writing
            partition.\\
            \param{$E$} & \inp & The number of bytes per array element.\\
		\end{apitabular}
        The data partition is defined by the array $(N_q)$ that must
be identically populated among all processes.

    \subsubsection{Write an array of variably sized elements}

	A more general function to write data arrays in parallel is \fscda{fwrite\_varray} that allows
	to write arrays with variable element size.
The encoding of the partition is accordingly more complex; see \secref{partition}.
The partition arguments $(N_q)$ and $(S_q)$ are again collective; for
transparency and non-redundancy we leave eventual allgather-type operations
to the caller.

        \apifunction{fwrite\_varray}{f, dbytes, $(N_q)_{<P}$,
                     $(E_i)_{C_{p}\leq i < C_{p+1}}$, $(S_q)_{<P}$,
                     \apitabreak
                     indirect, userstr, encode, err}
        \paragraph{Parameters}
        \begin{apitabular}
            \param{dbytes} & \inp
      & $N_p$ array elements of $E_i$ bytes per element and $S_p$ bytes overall.\\
            $(E_i)$ & \inp &
            The byte counts of the array elements local to this process.\\
            \param{$(S_q)$} & \inp & The array of byte counts per process.
        \end{apitabular}

    \subsection{Reading}
    \seclab{reading}

    For reading, we assume that a file valid according to our file format
    has been opened by \fscda{fopen} with mode \texttt{'r'} as specified in
    \secref{openclose}, which places the file cursor after the file header section.
    The remainder of the file is read one section at a time, where the type
    of section does not need to be known in advance.

    As for reading,
    we follow the terms introduced in
    \secref{partition} and \secref{conventions}.
    In particular, all parameters except
    \param{dbytes} and \param{$(E_i)$} are collective.
    This means that output parameters like \param{type}, $N$,
    $E$, and \param{userstr} are internally synchronized
    before returning.

    For reading a compressed file section
    as introduced in \secref{compression} it is sufficient
    to pass true for \param{decode} to \fscda{fread\_section\_header}.
    In this case, the file section data is decompressed on
    reading if its section header conforms to the compression convention.
    If no compressed data is encountered, the data is read as is.
    A value of false reads any encoded data raw.
    \begin{comment}
    Passing true for \param{decode} is always safe if
    the section header conforming to the compression
    convention is followed by data conforming to
    the compression convention or the section header
    does not satisfy the compression convention.
    Further details on the decode parameter can be found in the
    description of \fscda{fread\_section\_header} and in
    Table~\ref{tab:defineDecodes}.
    \end{comment}

	\subsubsection{Read a file section header}
	
	\fscda{fread\_section\_header} is a collective function
        to extract the upcoming file section type and its metadata.

	   \apifunction{fread\_section\_header}{f, type,
                    $N$, $E$,
                    \apitabreak
                    userstr, decode, err}

		\begin{apitabular}
			\param{type} & \outp & This is set to the file section type
            $t \in \{\I, \B, \A, \V\}$.
\\
			\param{$N$} & \outp & An integer that is set to the
            number of global array
            elements if $t \in \{\A, \V\}$.
            For $t \in \{\I, \B \}$ $N$ is set to 0.\\
			\param{$E$} & \outp & An integer that is set to the byte count
            of each array element for $t = \A$ and to the number
            of bytes in a data block for $t = \B$.
            Otherwise, $E$ is set to 0.\\
            \param{decode} & \inoutp & On input a Boolean to
            decide whether the file section shall possibly be
            interpreted as a compressed section.
            For true as input the file section
            is interpreted as a compressed file section if
            the type and user string of the first raw file section
            satisfiy the compression convention of \secref{compression}.
            If the compression convention
            is not satisfied the data is read raw.
            For false as input the data is read raw in any
            case. The output values depend on the input values and file
            contents as shown in Table~\ref{tab:defineDecodes}.
		\end{apitabular}
    \begin{table}
        \begin{center}
            \begin{tabular}{ccc}
                \toprule
                Input & \multicolumn{2}{c}{Output}\\
                & compression header & non-compression header\\
                \midrule
                $0$ & $0$ & $0$\\
                $1$ & $1$ & $0$\\
                \bottomrule
            \end{tabular}%
        \end{center}%
        \caption{Input and output for the \param{decode} argumment to
        the function \fscda{fread\_section\_header}.
        A compression header is encountered if the
        next file section contains a type and user string
        matching the compression convention described in
        \secref{compression}.
        If the input is false and a compression header is found,
        the compression is ignored and the data of this first section
        is read undecoded.
        If the input is true and a compression header is not found,
        the output value becomes false and we read the data as present in
        the file.}%
        \label{tab:defineDecodes}%
    \end{table}%

    For all four file section types we require further
    function calls, which must use parameters that are consistent with the
    output of \fscda{fread\_section\_header}.  In particular, this enables
    the user to write a query function that reads all file section
    headers but skips the data bytes to identify the structure of the file.
    The skipping of data bytes is described below for each file section
    type in turn.

    After determining the file section type and
    metadata using \fscda{fread\_section\_header}
    we are in the position to
    allocate further output variables and to read the data for $t \in \{\I, \B, \A\}$ or,
    respectively, the local array element sizes for $t = \V$.

    \subsubsection{Read data bytes of an inline section}
    The simplest file section provides 32 bytes of data.

    \apifunction{fread\_inline\_data}{f, dbytes, root, err}

    \paragraph{Parameters}
    \begin{apitabular}
            \param{dbytes} & \outp & 32 bytes memory on the process \param{root} that
            is filled
            with the data bytes of the inline section.
            The user can pass \NULL on \param{root} to skip the 32 bytes.
            The argument is ignored on non-\param{root} processes.
    \end{apitabular}

    \subsubsection{Read data bytes of a block section}

    We accomplish reading the data bytes of
    a block using another collective function:

        \apifunction{fread\_block\_data}{f, dbytes, $N$, root, err}
%\begin{samepage}
        \paragraph{Parameters}
        \begin{apitabular}
            \param{dbytes} & \outp & $N$ bytes memory on
            process \param{root}.
            For \NULL on the \param{root} process the data is skipped.
            On all other processes the argument is ignored.\\
            \param{$N$} & \inp & The byte count of the data bytes as
            retrieved from the preceding call to \fscda{fread\_section\_header}.
        \end{apitabular}
%\end{samepage}

    \subsubsection{Read data bytes of a fixed-size array section}

    As for the non-partitioned sections, the data of a fixed-size array
    section can be read after a call to \fscda{fread\_section\_header}.

        \apifunction{fread\_array\_data}{f, dbytes, $(N_q)_{<P}$, $E$,
            indirect, err}

        \paragraph{Parameters}
        \begin{apitabular}
            \param{dbytes} & \outp & $N_p$ array elements of $E$ bytes per element
            on each local process $p$.
            Passing \NULL on any process skips the array data on that process.\\
            \param{$(N_q)$} & \inp & The array element count per process.
            This array defines the reading
            partition and must satisfy $\sum_{q=0}^{P-1} N_q = N$
            as retrieved by the
            preceding call to \fscda{fread\_section\_header}.\\
            \param{$E$} & \inp & The number of bytes per array element as retrieved
            previously.
        \end{apitabular}

    \subsubsection{Read element sizes of a variable-size array}

    It remains to read a variable-size array.  For this file section
    type we need to first read the element sizes according to a given partition
    before we can read the actual array data.  For this purpose we provide
    \fscda{fread\_varray\_sizes}. In the following the notation is as in
    \eqnref{sp}.

        \apifunction{fread\_varray\_sizes}{f,
        $(E_i)_{C_{p}\leq i < C_{p+1}}$, $(N_q)_{<P}$, err}

        \paragraph{Parameters}
        \begin{apitabular}
            \param{$(E_i)$} & \outp & $N_p$ array elements of $8$ bytes each
            for an unsigned integer per element on the local process $p$,
            representing the byte counts of the process-local array elements.
            Passing \NULL on any process skips the size data on that process.\\
            \param{$(N_q)$} & \inp & The array element count per process.
            This array defines the reading
            partition and must satisfy $\sum_{q=0}^{P-1} N_q = N$
            as retrieved from the
            preceding call to \fscda{fread\_section\_header}.
        \end{apitabular}

    \subsubsection{Read data bytes of a variable-size array}

    Finally, one can read the actual variable-size array data using
    \fscda{fread\_varray\_data}. This function requires the array of
    byte counts per process on every calling process with \param{dbytes}
    not \NULL.  If this information is not known by context,
    it can be calculated from $(E_i)$ as retrieved from
    \fscda{fread\_varray\_sizes} applying \eqnref{sp}.

        \apifunction{fread\_varray\_data}{f, dbytes, $(N_q)_{<P}$,
        $(E_i)_{C_{p}\leq i < C_{p+1}}$,
        \apitabreak
        indirect, err}

%\begin{samepage}
        \paragraph{Parameters}
        \begin{apitabular}
            \param{\texttt{dbytes}} & \outp & $N_p$ array elements of $E_i$ bytes
            per element and $S_p$ bytes overall on
            the calling process $p$.
            For \NULL on any calling process $p$ the variable-size array data is skipped
            for this process.\\
            \param{$(N_q)$} & \inp & The array element count per process.
            This array defines the reading
            partition and must satisfy $\sum_{q=0}^{P-1} N_q = N$
            as retrieved from the
            preceding call to \fscda{fread\_section\_header}.\\
            \param{$(E_i)$} & \inp & The sizes of the process-local elements
            as retrieved from \fscda{fread\_varray\_sizes}.
            Ignored on every process where
            \NULL is passed for \param{dbytes}; must be consistent otherwise.\\
            \param{$(S_q)$} & \inp & The array of byte counts per process.
        \end{apitabular}
%\end{samepage}

    \subsection{Error management}
    \seclab{errorman}

    Dealing with file access is susceptible to errors that may occur even
    when using the API exactly as documented.
    Since the primary use of the exposed functionality is to support
    scientific computing workflows, say in providing
    simulation checkpointing, restart, and archival,
    and these jobs are often executed in a batch environment,
    file errors should never crash the simulation but instead
    allow for meaningful clean returns and exits.

    In order to give the user the chance to not abort
    and to
    receive a proper report of the concrete error, we always set an
    error code, which may then be reacted to by the user to potentially
    adjust e.\ g.\ the state of the file system, file locations and names,
    or to gain write permission.
    With the exceptions of blatantly violating the call conventions of a function,
    say passing \NULL for a mandatory parameter, which may trigger an assertion,
    we consider three groups of checked runtime errors, namely
    \begin{enumerate}
        \item corrupt file contents,
        \item file system errors, and
        \item semantically invalid input parameters or call sequence.
    \end{enumerate}
    The first group of errors includes invalid file section metadata and the
    second group any error reported by file system access functions.
    These file system dependent errors are in general a translated subset of
    the MPI I/O error classes \cite{Forum97} or \texttt{errno} \cite{ISO90}
    values depending on the availability of MPI I/O.
    Finally, the third group indicates that the user passed
    parameter(s) to an API function that have no legal meaning,
    or that multiple reading functions are improperly composed.

    \subsubsection{Retrieve an error string}

    All functions of the proposed API take an integer output parameter \texttt{err}
    that is set to the error code of the function call, or to 0 for no error.
    The code can, additionally, be translated to an error string using the
    following non-collective function:

        \apifunction{ferror\_string}{err, errorstr, strlen}

        \paragraph{Parameters}
        \begin{apitabular}
            \param{err} & \inp & The error code intended to be translated, including 0 for no error.\\
            \param{errorstr} & \outp & Set to a matching error string.\\
            \param{strlen} & \inoutp & The length of \param{errorstr}
                             $\{\text{in bytes on input}, \text{as actually output}\}$.
        \end{apitabular}

        \paragraph{Return}
        \apireturn{%
        0 in case of any valid input \param{err} and a negative value otherwise.}

    \bibliographystyle{ACM-Reference-Format}
    \bibliography{../bibtex/ccgo_new,../bibtex/carsten}


\begin{thebibliography}{21}

%%% the Web address.

\ifx \showCODEN    \undefined \def \showCODEN     #1{\unskip}     \fi
\ifx \showDOI      \undefined \def \showDOI       #1{#1}\fi
\ifx \showISBNx    \undefined \def \showISBNx     #1{\unskip}     \fi
\ifx \showISBNxiii \undefined \def \showISBNxiii  #1{\unskip}     \fi
\ifx \showISSN     \undefined \def \showISSN      #1{\unskip}     \fi
\ifx \showLCCN     \undefined \def \showLCCN      #1{\unskip}     \fi
\ifx \shownote     \undefined \def \shownote      #1{#1}          \fi
\ifx \showarticletitle \undefined \def \showarticletitle #1{#1}   \fi
\ifx \showURL      \undefined \def \showURL       {\relax}        \fi
% invisible to TeX
\providecommand\bibfield[2]{#2}
\providecommand\bibinfo[2]{#2}
\providecommand\natexlab[1]{#1}
\providecommand\showeprint[2][]{arXiv:#2}

\bibitem[Adler et~al\mbox{.}(2003)]%
        {AdlerBoutellBowlerEtAl03}
\bibfield{author}{\bibinfo{person}{Mark Adler}, \bibinfo{person}{Thomas
  Boutell}, \bibinfo{person}{John Bowler}, \bibinfo{person}{Christian
  Brunschen}, \bibinfo{person}{Adam~M. Costello}, \bibinfo{person}{Lee~Daniel
  Crocker}, \bibinfo{person}{Andreas Dilger}, \bibinfo{person}{Oliver Fromme},
  \bibinfo{person}{Jean loup Gailly}, \bibinfo{person}{Chris Herborth},
  \bibinfo{person}{Alex Jakulin}, \bibinfo{person}{Neal Kettler},
  \bibinfo{person}{Tom Lane}, \bibinfo{person}{Alexander Lehmann},
  \bibinfo{person}{Chris Lilley}, \bibinfo{person}{Dave Martindale},
  \bibinfo{person}{Owen Mortensen}, \bibinfo{person}{Keith~S. Pickens},
  \bibinfo{person}{Robert~P. Poole}, \bibinfo{person}{Glenn Randers-Pehrson},
  \bibinfo{person}{Greg Roelofs}, \bibinfo{person}{Willem van Schaik},
  \bibinfo{person}{Guy Schalnat}, \bibinfo{person}{Paul Schmidt},
  \bibinfo{person}{Michael Stokes}, \bibinfo{person}{Tim Wegner}, {and}
  \bibinfo{person}{Jeremy Wohl}.} \bibinfo{year}{2003}\natexlab{}.
\newblock \bibinfo{booktitle}{\emph{Portable Network Graphics ({PNG})
  Specification (Second Edition)}}.
\newblock \bibinfo{type}{{W3C} Recommendation}. \bibinfo{institution}{W3C}.
\newblock
\urldef\tempurl%
\url{https://www.w3.org/TR/2003/REC-PNG-20031110/}
\showURL{%
\tempurl}
\newblock
\shownote{Edited by David Duce}.


\bibitem[Boito et~al\mbox{.}(2018)]%
        {BoitoZanonInacioEtAl18}
\bibfield{author}{\bibinfo{person}{Francieli~Zanon Boito},
  \bibinfo{person}{Eduardo~C. Inacio}, \bibinfo{person}{Jean~Luca Bez},
  \bibinfo{person}{Philippe O.~A. Navaux}, \bibinfo{person}{Mario A.~R.
  Dantas}, {and} \bibinfo{person}{Yves Denneulin}.}
  \bibinfo{year}{2018}\natexlab{}.
\newblock \showarticletitle{A Checkpoint of Research on Parallel I/O for
  High-Performance Computing}.
\newblock \bibinfo{journal}{\emph{ACM Comput. Surv.}} \bibinfo{volume}{51},
  \bibinfo{number}{2}, Article \bibinfo{articleno}{23} (\bibinfo{date}{mar}
  \bibinfo{year}{2018}), \bibinfo{numpages}{35}~pages.
\newblock
\showISSN{0360-0300}
\urldef\tempurl%
\url{https://doi.org/10.1145/3152891}
\showDOI{\tempurl}


\bibitem[Braam and Schwan(2002)]%
        {BraamSchwan02}
\bibfield{author}{\bibinfo{person}{Peter~J Braam} {and} \bibinfo{person}{Philip
  Schwan}.} \bibinfo{year}{2002}\natexlab{}.
\newblock \showarticletitle{{L}ustre: The intergalactic file system}. In
  \bibinfo{booktitle}{\emph{Ottawa Linux Symposium}}.
  \bibinfo{publisher}{Ottawa Linux Symposium}, \bibinfo{address}{Ottawa,
  Ontario, Canada}, \bibinfo{pages}{50--54}.
\newblock


\bibitem[Burstedde(2010)]%
        {Burstedde23a}
\bibfield{author}{\bibinfo{person}{Carsten Burstedde}.}
  \bibinfo{year}{2010}\natexlab{}.
\newblock \bibinfo{title}{{\texttt{\upshape p4est}}: Parallel {AMR} on Forests
  of Octrees}.
\newblock
\newblock
\newblock
\shownote{\url{https://www.p4est.org/} (last accessed January 24th, 2023)}.


\bibitem[Burstedde et~al\mbox{.}(2014)]%
        {BursteddeCalhounMandliEtAl14}
\bibfield{author}{\bibinfo{person}{Carsten Burstedde}, \bibinfo{person}{Donna
  Calhoun}, \bibinfo{person}{Kyle~T. Mandli}, {and} \bibinfo{person}{Andy~R.
  Terrel}.} \bibinfo{year}{2014}\natexlab{}.
\newblock \showarticletitle{ForestClaw: Hybrid forest-of-octrees {AMR} for
  hyperbolic conservation laws}. In \bibinfo{booktitle}{\emph{Parallel
  Computing: Accelerating Computational Science and Engineering (CSE)}}
  \emph{(\bibinfo{series}{Advances in Parallel Computing},
  Vol.~\bibinfo{volume}{25})}, \bibfield{editor}{\bibinfo{person}{Michael
  Bader}, \bibinfo{person}{Arndt Bode}, \bibinfo{person}{Hans-Joachim
  Bungartz}, \bibinfo{person}{Michael Gerndt}, \bibinfo{person}{Gerhard~R.
  Joubert}, {and} \bibinfo{person}{Frans Peters}} (Eds.).
  \bibinfo{publisher}{IOS Press}, \bibinfo{address}{NLD},
  \bibinfo{pages}{253--262}.
\newblock
\urldef\tempurl%
\url{https://doi.org/10.3233/978-1-61499-381-0-253}
\showDOI{\tempurl}


\bibitem[Burstedde and Holke(2016)]%
        {BursteddeHolke16}
\bibfield{author}{\bibinfo{person}{Carsten Burstedde} {and}
  \bibinfo{person}{Johannes Holke}.} \bibinfo{year}{2016}\natexlab{}.
\newblock \showarticletitle{A Tetrahedral Space-Filling Curve for Nonconforming
  Adaptive Meshes}.
\newblock \bibinfo{journal}{\emph{SIAM Journal on Scientific Computing}}
  \bibinfo{volume}{38}, \bibinfo{number}{5} (\bibinfo{year}{2016}),
  \bibinfo{pages}{C471--C503}.
\newblock
\urldef\tempurl%
\url{https://doi.org/10.1137/15M1040049}
\showDOI{\tempurl}


\bibitem[Burstedde et~al\mbox{.}(2011)]%
        {BursteddeWilcoxGhattas11}
\bibfield{author}{\bibinfo{person}{Carsten Burstedde},
  \bibinfo{person}{Lucas~C. Wilcox}, {and} \bibinfo{person}{Omar Ghattas}.}
  \bibinfo{year}{2011}\natexlab{}.
\newblock \showarticletitle{{\texttt{\upshape p4est}}: Scalable Algorithms for
  Parallel Adaptive Mesh Refinement on Forests of Octrees}.
\newblock \bibinfo{journal}{\emph{SIAM Journal on Scientific Computing}}
  \bibinfo{volume}{33}, \bibinfo{number}{3} (\bibinfo{year}{2011}),
  \bibinfo{pages}{1103--1133}.
\newblock
\urldef\tempurl%
\url{https://doi.org/10.1137/100791634}
\showDOI{\tempurl}


\bibitem[Burstedde et~al\mbox{.}(2010)]%
        {BursteddeWilcoxIsaac23}
\bibfield{author}{\bibinfo{person}{Carsten Burstedde},
  \bibinfo{person}{Lucas~C. Wilcox}, {and} \bibinfo{person}{Tobin Issac}.}
  \bibinfo{year}{2010}\natexlab{}.
\newblock \bibinfo{title}{{\texttt{\upshape libsc}}: The "sc" auxiliary
  library}.
\newblock
\newblock
\newblock
\shownote{\url{https://github.com/cburstedde/libsc} (last accessed June 14th,
  2023)}.


\bibitem[Deutsch(1996)]%
        {Deutsch96}
\bibfield{author}{\bibinfo{person}{Peter Deutsch}.}
  \bibinfo{year}{1996}\natexlab{}.
\newblock \bibinfo{booktitle}{\emph{{DEFLATE} Compressed Data Format
  Specification version 1.3}}.
\newblock \bibinfo{type}{RFC} 1951. \bibinfo{institution}{RFC Editor}.
\newblock
\showISSN{2070-1721}
\urldef\tempurl%
\url{https://www.rfc-editor.org/rfc/rfc1951.txt}
\showURL{%
\tempurl}


\bibitem[Deutsch and Gailly(1996)]%
        {DeutschGailly96}
\bibfield{author}{\bibinfo{person}{Peter Deutsch} {and}
  \bibinfo{person}{Jean-Loup Gailly}.} \bibinfo{year}{1996}\natexlab{}.
\newblock \bibinfo{booktitle}{\emph{{ZLIB} Compressed Data Format Specification
  version 3.3}}.
\newblock \bibinfo{type}{RFC} 1950. \bibinfo{institution}{RFC Editor}.
\newblock
\showISSN{2070-1721}
\urldef\tempurl%
\url{https://www.rfc-editor.org/rfc/rfc1950.txt}
\showURL{%
\tempurl}


\bibitem[Gailly and Adler(2023)]%
        {GaillyAdler23}
\bibfield{author}{\bibinfo{person}{Jean-Loup Gailly} {and}
  \bibinfo{person}{Mark Adler}.} \bibinfo{year}{2023}\natexlab{}.
\newblock \bibinfo{title}{zlib repository}.
\newblock
\newblock
\newblock
\shownote{\url{https://github.com/madler/zlib} (last accessed June 6th, 2023)}.


\bibitem[Holke et~al\mbox{.}(2023)]%
        {HolkeBursteddeKnappEtAl23}
\bibfield{author}{\bibinfo{person}{Johannes Holke}, \bibinfo{person}{Carsten
  Burstedde}, \bibinfo{person}{David Knapp}, \bibinfo{person}{Lukas Dreyer},
  \bibinfo{person}{Sandro Elsweijer}, \bibinfo{person}{Veli {\"U}nl{\"u}},
  \bibinfo{person}{Johannes Markert}, \bibinfo{person}{Ioannis Lilikakis},
  \bibinfo{person}{Niklas B{\"o}ing}, \bibinfo{person}{Prasanna Ponnusamy},
  {and} \bibinfo{person}{Achim Basermann}.} \bibinfo{year}{2023}\natexlab{}.
\newblock \showarticletitle{\texttt{t8code} v. 1.0 - Modular Adaptive Mesh
  Refinement in the Exascale Era}. In \bibinfo{booktitle}{\emph{{SIAM}
  International Meshing Round Table 2023}}. \bibinfo{publisher}{{SIAM}},
  \bibinfo{address}{Amsterdam, NL}, \bibinfo{numpages}{5}~pages.
\newblock
\urldef\tempurl%
\url{https://elib.dlr.de/194377/}
\showURL{%
\tempurl}


\bibitem[Inc.(2010)]%
        {Kitware10}
\bibfield{author}{\bibinfo{person}{Kitware Inc.}}
  \bibinfo{year}{2010}\natexlab{}.
\newblock \bibinfo{booktitle}{\emph{The {VTK} User's Guide}
  (\bibinfo{edition}{11th} ed.)}.
\newblock \bibinfo{publisher}{Kitware}, \bibinfo{address}{Clifton Park, NY}.
\newblock
\showISBNx{9781930934238}
\urldef\tempurl%
\url{https://vtk.org/wp-content/uploads/2021/08/VTKUsersGuide.pdf}
\showURL{%
\tempurl}


\bibitem[ISO(1990)]%
        {ISO90}
\bibfield{author}{\bibinfo{person}{ISO}.} \bibinfo{year}{1990}\natexlab{}.
\newblock \bibinfo{booktitle}{\emph{ISO C Standard C89/C90}}.
\newblock \bibinfo{type}{{T}echnical {R}eport}. \bibinfo{institution}{ISO}.
\newblock
\newblock
\shownote{ISO/IEC 9899:1990}.


\bibitem[Jain et~al\mbox{.}(2021)]%
        {JainWeideChawdharyEtAl21}
\bibfield{author}{\bibinfo{person}{Rajeev Jain}, \bibinfo{person}{Klaus Weide},
  \bibinfo{person}{Saurabh Chawdhary}, {and} \bibinfo{person}{Thomas
  Klostermann}.} \bibinfo{year}{2021}\natexlab{}.
\newblock \bibinfo{title}{Checkpoint/Restart for Lagrangian particle mesh with
  AMR in community code FLASH-X}.
\newblock
\newblock
\urldef\tempurl%
\url{http://arxiv.org/abs/2103.04267}
\showURL{%
\tempurl}


\bibitem[Li et~al\mbox{.}(2003)]%
        {LiLiaoChoudharyEtAl03}
\bibfield{author}{\bibinfo{person}{Jianwei Li}, \bibinfo{person}{Wei-keng
  Liao}, \bibinfo{person}{Alok Choudhary}, \bibinfo{person}{Robert Ross},
  \bibinfo{person}{Rajeev Thakur}, \bibinfo{person}{William Gropp},
  \bibinfo{person}{Rob Latham}, \bibinfo{person}{Andrew Siegel},
  \bibinfo{person}{Brad Gallagher}, {and} \bibinfo{person}{Michael Zingale}.}
  \bibinfo{year}{2003}\natexlab{}.
\newblock \showarticletitle{Parallel {NetCDF}: A High-Performance Scientific
  {I/O} Interface}. In \bibinfo{booktitle}{\emph{Proceedings of the 2003
  ACM/IEEE Conference on Supercomputing}} (Phoenix, AZ, USA)
  \emph{(\bibinfo{series}{SC '03})}. \bibinfo{publisher}{Association for
  Computing Machinery}, \bibinfo{address}{New York, NY, USA},
  \bibinfo{pages}{39}.
\newblock
\showISBNx{1581136951}
\urldef\tempurl%
\url{https://doi.org/10.1145/1048935.1050189}
\showDOI{\tempurl}


\bibitem[{Message Passing Interface Forum}(1997)]%
        {Forum97}
\bibfield{author}{\bibinfo{person}{{Message Passing Interface Forum}}.}
  \bibinfo{year}{1997}\natexlab{}.
\newblock \bibinfo{title}{{MPI}: A Message-Passing Interface Standard, Version
  2.0}.
\newblock
\newblock
\newblock
\shownote{\url{https://www.mpi-forum.org/docs/mpi-2.0/mpi-20.ps}, Last accessed
  on May 4th, 2023}.


\bibitem[Rew et~al\mbox{.}(1997)]%
        {RewDavisEmmersonEtAl97}
\bibfield{author}{\bibinfo{person}{Russ Rew}, \bibinfo{person}{Glenn Davis},
  \bibinfo{person}{Steve Emmerson}, {and} \bibinfo{person}{Harvey~L. Davies}.}
  \bibinfo{year}{1997}\natexlab{}.
\newblock \bibinfo{booktitle}{\emph{{NetCDF} User's Guide - An Interface for
  Self-Describing, Portable Data, Version 3}}.
\newblock \bibinfo{publisher}{Unidata Program Center},
  \bibinfo{address}{Boulder, CO}.
\newblock


\bibitem[Strafella and Chapon(2022)]%
        {StrafellaChapon22}
\bibfield{author}{\bibinfo{person}{Loic Strafella} {and}
  \bibinfo{person}{Damien Chapon}.} \bibinfo{year}{2022}\natexlab{}.
\newblock \showarticletitle{{LightAMR} format standard and lossless compression
  algorithms for adaptive mesh refinement grids: {RAMSES} use case}.
\newblock \bibinfo{journal}{\emph{J. Comput. Phys.}}  \bibinfo{volume}{470}
  (\bibinfo{year}{2022}), \bibinfo{pages}{111577}.
\newblock
\showISSN{0021-9991}
\urldef\tempurl%
\url{https://doi.org/10.1016/j.jcp.2022.111577}
\showDOI{\tempurl}


\bibitem[{The HDF Group}(2023)]%
        {HDF23}
\bibfield{author}{\bibinfo{person}{{The HDF Group}}.}
  \bibinfo{year}{1997-2023}\natexlab{}.
\newblock \bibinfo{title}{{Hierarchical Data Format, version 5}}.
\newblock
\newblock
\newblock
\shownote{\url{https://www.hdfgroup.org/HDF5/}}.


\bibitem[{The HDF Group}(2019)]%
        {HDF19}
\bibfield{author}{\bibinfo{person}{{The HDF Group}}.}
  \bibinfo{year}{2019}\natexlab{}.
\newblock \bibinfo{title}{{HDF}5 File Format Specification Version 3.0}.
\newblock
\newblock
\newblock
\shownote{\url{https://portal.hdfgroup.org/download/attachments/52627880/HDF5_File_Format_Specification_Version-3.0.pdf?api=v2},
  Last accessed on May 3rd, 2023}.


\end{thebibliography}
\end{document}